# CoRoT Measures Solar-Like Oscillations and Granulation in Stars Hotter Than the Sun

Eric Michel,[1]* Annie Baglin,[1] Michel Auvergne,[1] Claude Catala,[1] Reza Samadi,[1] Frédéric Baudin,[2] Thierry Appourchaux,[2] Caroline Barban,[1] Werner W. Weiss,[3] Gabrielle Berthomieu,[4] Patrick Boumier,[2] Marc-Antoine Dupret,[1] Rafael A. Garcia,[5] Malcolm Fridlund,[6] Rafael Garrido,[7] Marie-Jo Goupil,[1] Hans Kjeldsen,[8] Yveline Lebreton,[9] Benoît Mosser,[1] Arlette Grotsch-Noels,[10] Eduardo Janot-Pacheco,[11] Janine Provost,[4] Ian W. Roxburgh,[12,1] Anne Thoul,[10] Thierry Toutain,[13] Didier Tiphène,[1] Sylvaine Turck-Chieze,[5] Sylvie D. Vauclair,[14] Gérard P. Vauclair,[14] Conny Aerts,[15] Georges Alecian,[16] Jérôme Ballot,[17] Stéphane Charpinet,[14] Anne-Marie Hubert,[9] François Lignières,[14] Philippe Mathias,[18] Mario J. P. F. G. Monteiro,[19] Coralie Neiner,[9] Ennio Poretti,[20] José Renan de Medeiros,[21] Ignasi Ribas,[22] Michel L. Rieutord,[14] Teodoro Roca Cortés,[23] Konstanze Zwintz[3]

Oscillations of the Sun have been used to understand its interior structure. The extension of similar studies to more distant stars has raised many difficulties despite the strong efforts of the international community over the past decades. The CoRoT (Convection Rotation and Planetary Transits) satellite, launched in December 2006, has now measured oscillations and the stellar granulation signature in three main sequence stars that are noticeably hotter than the sun. The oscillation amplitudes are about 1.5 times as large as those in the Sun; the stellar granulation is up to three times as high. The stellar amplitudes are about 25% below the theoretic values, providing a measurement of the nonadiabaticity of the process ruling the oscillations in the outer layers of the stars.

The discovery of global oscillations in the Sun (*1*, *2*) opened the way to solar seismology, that is, to sounding the Sun's interior, measuring, for instance, the depth of its convection zone and its rotation at different depths and latitudes (*3*). High-precision photometry from space has long been considered the best way to extend these techniques to other main sequence stars of moderate mass where such oscillations are expected. However, the first attempts were ambiguous (*4*, *5*), casting some doubt on the theoretical estimates of intrinsic amplitudes and questioning to what extent the oscillations might be hidden by stellar granulation. We present here the detection of solar-like oscillations in three stars observed by the CoRoT (Convection Rotation and Planetary Transits) (*6*) space mission, and we characterize their amplitudes and the granulation signature.

Detecting and measuring solar-like oscillations in main sequence stars other than the Sun is challenging. Tracking the variations in the light from a star to one part per million (ppm) requires high accuracy on individual measurements. It also requires long uninterrupted sequences of observations to enhance the statistics of the measurements without being polluted by the spurious frequency components induced by data gaps. Solar-like oscillations have been detected from the ground in radial velocity in several stars (Fig. 1). However, ground-based observations are hampered by diurnal interruptions, weather instabilities, and the annual motion of the Earth. As a result, all existing data sets suffer more or less severely from a limited time base and large gaps in the data, which hamper the measurement of mode characteristics. In addition, radial velocity observations are strongly biased toward low-effective temperature stars (and slow rotators), for they require many narrow spectral lines, and toward subgiant and giant stars, which show oscillations of the same nature as the Sun and other main sequence stars but with larger intrinsic amplitudes. On the other hand, photometric detection of solar-like oscillations has not been possible from the ground because of the higher sensitivity to atmospheric scintillation, and the previous space projects detected only power excess so far [for Procyon and beta Hydri with WIRE (Wide-Field Infrared Explorer) (*7*, *8*) and eta Boo with MOST (Microvariabilité et Oscillations

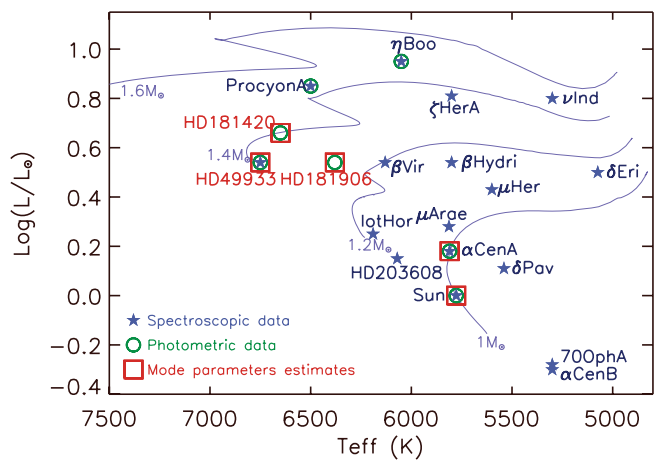

**Fig. 1.** HR diagram featuring stars for which mode structure has been observed in photometry (red squares), a power excess has been detected in photometry (green circles), and a detection has been performed in radial velocity (blue stars). Stellar evolutionary tracks are taken from (*20*), for solar chemical composition. Red giant pulsators (~6 objects) are out of the upper right corner of the figure.

[1]Laboratoire d'Etudes Spatiales et d'Instrumentation Astrophysique (LESIA), Observatoire de Paris, CNRS (UMR 8109)-Université Paris 6 Pierre et Marie Curie-Université Paris 7 Denis Diderot, Place Jules Janssen, F- 92195 Meudon, France. [2]Institut d'Astrophysique Spatiale (IAS), Université Paris-Sud 11, CNRS (UMR 8617), Bâtiment 121, F-91405 Orsay, France. [3]Institute for Astronomy, University of Vienna, Tuerkenschanzstrasse 17, A-1180 Vienna, Austria. [4]Laboratoire Cassiopée, Observatoire de la Côte d'Azur, CNRS (UMR 6202), BP 4229, F-06304 Nice Cedex 04, France. [5]Laboratoire Astrophysique Interactions Multi-échelles (AIM), Commissariat à l'Energie Atomique/DSM (Direction des Sciences de la Matière)-CNRS-Université Paris 7 Denis Diderot, CEA, IRFU (Institut de Recherche sur les Lois Fondamentales de l'Univers), SAp (Service d'Astrophysique), F-91191 Gif-sur-Yvette Cedex, France. [6]Astrophysics Mission Division, Research and Scientific Support Department, European Space Agency (ESA), European Space Research and Technology Center, SCI-SA, Post Office Box 299, Keplerlaan 1 NL-2200AG, Noordwijk, Netherlands. [7]Instituto de Astrofísica de Andalucía (CSIC) C/ Camino Bajo de Huétor, 50 E-18008 Granada, Spain. [8]Danish AstroSeismology Centre (DASC) Institut for Fysik og Astronomi Aarhus Universitet Bygning 1520, Ny Munkegade DK-8000 Aarhus C, Denmark. [9]Laboratoire Galaxies, Etoiles, Physique et Instrumentation (GEPI), Observatoire de Paris, CNRS (UMR 8111), Place Jules Janssen, F-92195 Meudon, France. [10]Institut d'Astrophysique et de Géophysique Université de Liège, Allée du 6 Août 17, B-4000 Liège, Belgique. [11]Instituto de Astronomia, Geofisica e Ciencias Atmosfericas, Rua do Matao, 1226/05508-090 Sao Paulo, Brasil. [12]Queen Mary University of London, Mile End Road, London E1 4NS, UK. [13]School of Physics and Astronomy, University of Birmingham, Edgbaston B15 2TT, UK. [14]Laboratoire d'Astrophysique de Toulouse-Tarbes, Université de Toulouse, CNRS (UMR 5572), 14 Avenue Edouard Belin, F-31400 Toulouse, France. [15]Instituut voor Sterrenkunde, Departement Natuurkunde en Sterrenkunde, Katholieke Universiteit Leuven, Celestijnenlaan 200 D, B -3001 Leuven, Belgium. [16]Laboratoire Univers et Théories (LUTH), Observatoire de Paris, CNRS (UMR 8102), Université Paris 7 Denis Diderot, 5 Place Jules Janssen, F-92190 Meudon, France. [17]Max Planck Institut für Astrophysik, Karl-Schwarzschild-Strasse 1, Postfach 1317, D-85741 Garching, Germany. [18]Laboratoire Hippolyte Fizeau, Observatoire de la Côte d'Azur, CNRS (UMR 6525), Université Nice Sophia-Antipolis, Campus Valrose, F-06108 Nice Cedex 2, France. [19]Centro de Astrofisica da Universidade do Porto, Rua das Estrelas, 4150-762 Porto, Portugal. [20]Istituto Nazionale di Astrofisica-Osservatorio Astronomico di Brera Via Emilio Bianchi 46, 23807 Merate (LC), Italy. [21]Departamento de Física, Universidade Federal do Rio Grande do Norte, 59072-970, Natal RN, Brasil. [22]Institut de Ciencies de l'Espai (CSIC-IEEC) Campus UAB Facultat de Ciències, Torre C5-parell, 2a pl 08193 Bellaterra, Spain. [23]Instituto de Astrofísica de Canarias, and Departamento de Astrofísica, Universidad de La Laguna, 38207 La Laguna, Tenerife, Spain.

*To whom correspondence should be addressed. E-mail: Eric.Michel@obspm.fr

**Table 1.** Parameters obtained in the present analysis, with standard deviation estimates.

| Star | $A_{bol}$ ($l = 0$)(ppm) | $B_{bol}$ (ppm$^2$/μHz) | C (s) | Δ (μHz) |
|---|---|---|---|---|
| HD 49933 | 4.02 ± 0.57 | 1.97 ± 0.53 | 1967 ± 431 | 86 ± 2 |
| HD 181420 | 3.82 ± 0.40 | 2.41 ± 0.31 | 1936 ± 206 | 77 ± 2 |
| HD 181906 | 3.26 ± 0.42 | 1.12 ± 0.20 | 1650 ± 0276 | 88 ± 2 |
| Sun PMO6 | 2.39 ± 0.17 | 0.85 ± 0.06 | 1440 ± 86 | 135 ± 2 |

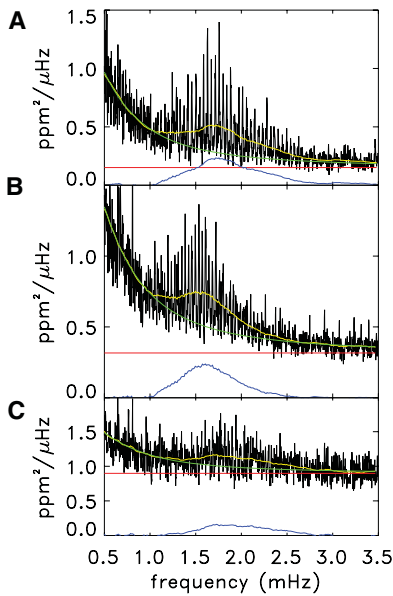

**Fig. 2.** Instrumental power spectral density. (**A**) For HD49933; a moving mean is applied with a 4-μHz boxcar (black); yellow curve: same spectrum highly smoothed (4 times Δ boxcar); green curve: mean level of the granulation + white noise components; red curve: mean white noise component level alone; blue curve: oscillation mean power density contribution alone. (**B**) Same for HD181420. (**C**) Same for HD181906.

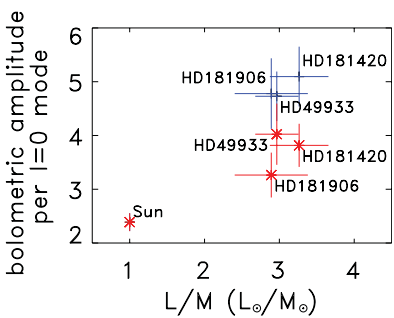

**Fig. 3.** Maximum bolometric amplitudes per radial mode measured (red) for HD49933, HD181420, HD181906, and for the Sun. Theoretical values are also given (blue). Error bars on amplitudes are standard deviation estimates associated with the accuracy of the measurements (red), and with the error estimate on $T_{eff}$ (blue).

Stellaires) (*9*)]. For alpha Cen A, WIRE (*10*) detected the characteristic comblike pattern of the oscillations, which could be analyzed with the help of complementary velocity data (*11*). However, alpha Cen A is very close to the Sun in terms of its global characteristics. The results here are based on light curves obtained with CoRoT over 60 days for HD49933 and 156 days for HD181420 and HD181906, three main sequence F stars noticeably hotter than the Sun (Fig. 1 and table S1).

The CoRoT satellite was launched on December 2006 in an inertial polar orbit at an altitude of 897 km. The instrument is fed by a 27-cm diameter telescope. During each run, it simultaneously provides light curves (variations in stellar flux with time) from 10 bright stars ($5.5 < m_V < 9.5$) dedicated to seismic studies, while 12,000 fainter stars ($11.5 < m_V < 15.5$) are monitored to search for transits due to planets (*6*). The sampling rate is 1 s for an integration time of 0.794 s. Pointing stability reaches a precision of 0.15″ root mean square. The duty cycle was higher than 93%; the missing data correspond essentially to the time spent in the South Atlantic magnetic anomaly where the perturbations due to energetic particles have not, as yet, been effectively corrected. These gaps, about eight per day, from 5 to 15 min each, have been linearly interpolated (with a 2000-s boxcar on each side of the gap to prevent the introduction of any spurious high frequencies) before we computed the Fourier power spectra, to minimize the aliases of the low-frequency components due to the window. We used synthetic spectra to check that this procedure has no noticeable influence on the measured mean values.

For each of the three stars, the Fourier power density spectra (Fig. 2) show three components that can be understood as (i) a flat white-noise component essentially due to photon counting noise, (ii) a stellar background component (essentially granulation in this frequency domain) following a Lorentzian profile $B/[1+(Cv)^2]$ as suggested in (*12*), and (iii) the stellar oscillation spectrum with its comblike pattern characterized by the large separation Δ (*13*).

Although dedicated analyses are under way to extract individual mode frequencies and profiles for each star, we measure here the contributions of these three components. We follow the method proposed in (*14*) and illustrated in Fig. 2, and we convert these instrumental values into intrinsic bolometric maximum amplitude per radial mode [$A_{bol}(l = 0)$] and bolometric maximum power spectral density $B_{bol}$ (*15*). We apply the same analysis to the solar SOHO/VIRGO/PMO6 (Solar and Heliospheric Observatory/

Variability of Solar Irradiance and Gravity Oscillations) data (*16*). The amplitudes of the three stars are larger than in the Sun by a factor of ~1.5 (Fig. 3).

Theoretical predictions suggest that velocity amplitudes follow a scaling law in $(L/M)^\alpha$ with α ~ 0.7 (*L* and *M* standing for luminosity and mass), in broad agreement with the existing velocity measurements (*17*). In the adiabatic approximation (*18*), this would give photometric amplitudes scaling as $(L/M)^\alpha (T_{eff})^{1/2}$, where $T_{eff}$ is effective temperature. As shown in Fig. 3 (see also Table 1), the measured values for the three stars are of the same order but significantly lower (by 24 ± 8% globally) than the theoretical values. The measurement of this systematic departure from the adiabatic case, which is not observed in velocity, tells us about the exchange of energy between convection and oscillations in the outer part of the convection zone. This process is responsible for the existence, and the specific amplitudes and lifetimes, of the oscillations. Both radial velocity and photometry measurements are sensitive to the oscillation momentum induced by this energy exchange; the photometric amplitudes are in addition more sensitive to the details of this process, via radiation-matter interaction. These measurements offer the possibility of testing theoretical models of the nonadiabatic effects of the processes governing the oscillations and illustrate the complementary interest of photometry and radial velocity measurements (when they are possible), which probe the oscillations differently.

The spectral signature of granulation is expected to reveal time scales and distance scales characteristic of the convection process in different stars (*12*, *19*). Our data show (fig. S1 and Table 1) that (i) the maximum bolometric power density ($B_{bol}$), associated with the number of eddies seen at the stellar surface and the border/center contrast of the granules, is higher for the three stars than for the Sun by a factor up to 3; and (ii) the characteristic time scale for granulation (C) associated with the eddy turnover time increases slightly with $T_{eff}$ (up to 30% higher than the Sun).

Supporting Online Material:

# CoRoT measurements of solar-like oscillations and granulation in stars hotter than the Sun

Eric Michel[1*], Annie Baglin[1], Michel Auvergne[1], Claude Catala[1], Reza Samadi[1], Frédéric Baudin[2], Thierry Appourchaux[2], Caroline Barban[1], Werner W. Weiss[3], Gabrielle Berthomieu[4], Patrick Boumier[2], Marc-Antoine Dupret[1], Rafael A. Garcia[5], Malcolm Fridlund[6], Rafael Garrido[7], Marie-Jo Goupil[1], Hans Kjeldsen[8], Yveline Lebreton[9], Benoît Mosser[1], Arlette Grotsch-Noels[10], Eduardo Janot-Pacheco[11], Janine Provost[4], Ian W Roxburgh[12,1], Anne Thoul[10], Thierry Toutain[13], Didier Tiphène[1], Sylvaine Turck-Chieze[5], Sylvie D. Vauclair[14], Gérard P. Vauclair[14], Conny Aerts[15], Georges Alecian[16], Jérôme Ballot[17], Stéphane Charpinet[14], Anne-Marie Hubert[9], François Lignières[14], Philippe Mathias[18], Mario JPFG Monteiro[19], Coralie Neiner[9], Ennio Poretti[20], José Renan de Medeiros[21], Ignasi Ribas[22], Michel L. Rieutord[14], Teodoro Roca Cortés[23] & Konstanze Zwintz[3]

In our analysis, as suggested in (*S1*), the power spectra are smoothed with a boxcar, taken here to be four times the large separation Δ (*S2*) estimated from the autocorrelation of each oscillation spectrum. Then, an estimate of the first two components (white noise and stellar background) is obtained by a least squares fit of the spectrum outside the domain where the oscillation signal is seen. After subtraction of these two components, we isolate that due to stellar oscillations ($P_o$). Instrumental noise is neglected here, since the same analysis applied to bright hot stars showing smaller or no significant granulation signature allows setting a higher limit for instrumental noise, at least 10 times below the values measured for the three solar-like pulsators.

In order to compare with theoretical estimates and with measurements on the Sun, it is convenient, following (*S1*) [see also (*S3*)] to convert these instrumental power spectral densities into intrinsic bolometric amplitudes per radial mode: $A_{bol}(l=0)=4(2P_o)^{1/2}\Delta/R_o$, and bolometric power spectral density of the granulation $B_{bol}/(1+(C\nu)^2)$ (see Fig. S1), where $B_{bol}=B/R_g$, and $R_o$ and $R_g$ are the instrumental response functions for CoRoT (*S3*).

**Fig. S1.** Bolometric power spectral density of granulation measured for HD49933 (red), HD181420 (green), HD181906 (blue) and for reference for the Sun (black). Error bars are standard deviation estimates resulting from a least squares fit.

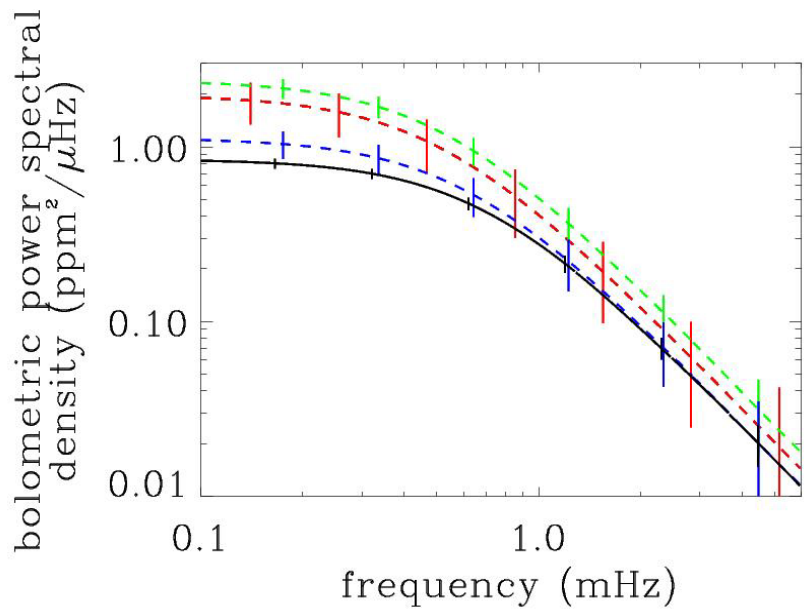

**Table S1.** Fundamental parameters of the stars considered here, with standard deviation estimates. Effective Temperatures and chemical contents have been derived via spectroscopic detailed analysis as described in (*S4*). Luminosities are computed using Hipparcos parallaxes (*S5*) and bolometric corrections from (*S6*). Masses are obtained by comparison with an extensive grid of stellar models described in (*S7*).

| Star | $T_{eff}$ (K) | [Fe/H] | Log(L/Lsun) | M/Msun |
|---|---|---|---|---|
| HD 49933 | 6750±60 | -0.4±0.1 | 0.54±0.02 | 1.17±0.1 |
| HD 181420 | 6650±60 | -0.04±0.1 | 0.66±0.04 | 1.4±0.1 |
| HD 181906 | 6380±60 | -0.14±0.1 | 0.54±0.06 | 1.2±0.1 |

**References and Notes**

S1. H. Kjeldsen *et al.* Solar-like Oscillations in α Centauri B. *Astrophys. J.* **635**, 1281-1290 (2005).

S2. The large separation (Δ) refers to the first order regular spacing in frequency between consecutive overtone eigenfrequencies, responsible for the characteristic comb-like pattern of the oscillation spectrum.

S3. E. Michel *et al.* Intrinsic stellar oscillation amplitude and granulation power density measured in photometry. *submitted to Astron, Astrophys.*(2008) http://arxiv.org/abs/0809.1078